# Rapid Size-Controlled Synthesis of Dextran-Coated, $^{64}$Cu-Doped Iron Oxide Nanoparticles

*Ray M. Wong[1], Dustin Gilbert[2], Kai Liu,[2] Angelique Y. Louie[3]**

Department of Chemistry[1], Physics[2], Biomedical Engineering[3], University of California Davis, California, 95616, USA

*corresponding author: aylouie@ucdavis.edu

Running title: Microwave Synthesis of DIO/Cu

***Abstract:*** Research into developing dual modality probes enabled for magnetic resonance imaging (MRI) and positron emission tomography (PET) has been on the rise recently due to the potential to combine the high resolution of MRI and the high sensitivity of PET. Current synthesis techniques for developing multimodal probes is largely hindered in part by prolonged reaction times during radioisotope incorporation – leading to a weakening of the radioactivity. Along with a time-efficient synthesis, the resulting products must fit within a critical size range (between 20-100nm) to increase blood retention time. In this work, we describe a novel, rapid, microwave-based synthesis technique to grow dextran-coated iron oxide nanoparticles doped with copper (DIO/Cu). Traditional methods for coprecipitation of dextran-coated iron oxide nanoparticles require refluxing for 2 hours and result in approximately 50 nm diameter particles. We demonstrate that microwave synthesis can produce 50 nm nanoparticles with 5 minutes of heating. We discuss the various parameters used in the microwave synthesis protocol to vary the size distribution of DIO/Cu, and demonstrate the successful incorporation of $^{64}$Cu into these particles with the aim of future use for dual-mode MR/PET imaging.

## Introduction

The ability to combine different imaging modalities has become an exciting area of medical research, due in part to recent breakthroughs in fused instrumentation, allowing simultaneous imaging with more than one technique. For example, in the past decade PET/Computed Tomography (CT) has been one of the most rapid growth areas in clinical imaging.[1] CT provides anatomical context for the PET images, which can otherwise be difficult to interpret.[2] Current attention has turned to fused PET/MRI instruments,[3] two modalities which had previously been presumed to be incompatible.[4] The marriage of these modalities provides images that combine the high resolution, anatomical imaging strengths of MRI, which has greater soft tissue contrast than CT, with ability of PET to detect biomarkers at high sensitivity, and monitor bio pathways. The development of probes detectable by both modalities can be advantageous, allowing image co-registration, and mapping of biomarkers across resolution scales. Only recently have PET/MRI dual-mode probes emerged in the literature.[3] Radiolabeled probes built upon iron oxide nanoparticles, a clinical MRI agent, have been studied by our lab[5, 6, 7] and others.[8, 9] A common approach to generate radiolabeled nanoparticles is to attach $^{64}Cu$ through chelation onto the surface.[10] However, the chelation and purification process is relatively slow and the stability of chelated $^{64}Cu$ *in vivo* is often uncertain.[11] It would be desirable to develop a synthesis method which could rapidly produce labeled particles that are stable *in vivo* – avoiding the ambiguity of whether free copper or labeled probe is providing the PET signal.

Microwave heating is known to reduce reaction times and improve synthesis yield by increasing the homogeneity of heating.[12] The decreased reaction time offered by a microwave approach is extremely important to the synthesis and application of a dual modality probe. Traditional iron oxide synthesis methods, such as coprecipitation, typically require a long (2h+) reflux[6] which represents a significant portion of the 12.7 h half life of $^{64}Cu$ radiolabel.[13] Microwave synthesis has been demonstrated in the growth of $\gamma$-$Fe_2O_3$,[14] Au and Pd nanoparticles,[15] and uncoated iron oxide nanoparticles.[16, 17] However, for successful medical application, iron oxide particles must be coated for hydrostatic stability, protection from degradation, and biocompatibility.

In this work we develop a one-step technique to synthesize dextran-coated iron oxide nanoparticles (DIO) doped with copper (DIO/Cu) and copper-64 – a PET contrast agent – using microwave assisted heating. This method incorporates a radiolabel as part of the integral structure of the iron oxide nanoparticles. By doing this we predict that the copper would be more stable against release from nanoparticles *in vivo* compared to surface chelated copper.  In addition, the release of free copper may be more easily distinguished than the loss of chelated copper in that particle breakdown would be required for the release of copper doped into the iron oxide particle core, which would result in loss of MRI signal. Nonradioactive copper was used for proof-of-principle studies, followed by incorporation of radioactive $^{64}Cu$ to validate the method. Further, we vary the synthesis conditions and observe variations in particle size, and also evaluate the magnetic properties of these particles. The microwave-assisted method significantly decreases synthesis time, which reduces the loss of radioactivity, and forms a more stable particle, and also is far simpler than other alternative methods of fabricating radio-labeled nanoparticles.

**Experimental Section**

We have previously reported size-controlled syntheses of dextran coated iron oxide nanoparticles[6] and used this platform for our DIO/Cu synthesis. Reagents were obtained from either Sigma-Aldrich or Fisher Scientific and were used directly, unless otherwise noted. A Millipore Milli-Q purifier (18.0 MΩ cm, Barnstead) was used to purify water.

**Preparation of reduced dextran.** Reduced dextran was prepared by modifying a reported method[7, 18]. 10 g of dextran (MW/MN = 10,000 Da) was dissolved in 100 ml of nanopure water at 25 $^{o}$C, followed by the addition of 1 g of sodium borohydride. The mixture was stirred for 24 hours. The pH was then brought to pH 5 with 6N HCl and the mixture was put into membrane dialysis bag with a filter size of 15,000 Da (Spectra, California, USA) for 3 days (8-10 changes of water). The product was then lyophilized and stored at 4°C.

**Synthesis of DIO and DIO/Cu nanoparticles.** DIO/Cu was prepared by first degassing nanopure water by argon bubbling. 0.02 mmol of $FeCl_3 \cdot 6H_2O$ was combined with 0.007 mmol of reduced dextran and dissolved in 3.7 ml of degassed $H_2O$. This solution was stirred for 40 min under an argon blanket at 4°C. $FeCl_2 \cdot 4H_2O$ (0.137 mmol) and $CuCl_2$ (0.0137 mmol) were subsequently prepared by dissolving each in 1.5 ml of degassed $H_2O$ and stored at 4°C for 20 minutes. The prepared $FeCl_2 \cdot 4H_2O$ and $CuCl_2$ were quickly added to the $FeCl_3 \cdot 6H_2O$ and reduced dextran mixture, and stirred for 5 minutes. 500 μl of $NH_4OH$ was then added at a rate of 0.1 ml every 5 seconds. The mixture was then loaded into a 30ml Activent microwave tube and heated in the microwave

reactor – Automated Microwave Explorer/Discover Hybrid-12 (CEM Corporation, USA) – under various times and power. The resulting nanoparticles were evaluated after microwave heating for 5, 10, and 15 minutes while maintaining a maximum power output of 300W. For evaluation of the effect of power level on the synthesis, reaction time was fixed at 10 minutes and power outputs set to 150, 200, 250, and 300W (300W was the maximum power setting available). The microwave system typically is closed, with an emergency release of pressure only above a maximum threshold. Dextran coated iron oxide (DIO) was synthesized using the above synthesis method in the absence of copper.

**Synthesis of Cu-64 nanoparticles**. Using a similar reaction scheme to the above paragraph, 0.0137mmol of $CuCl_2$ was replaced with 1 mCi of $Cu^{64}Cl_2$ and the microwave time was set to 1 minute. Purification was performed with a 10kDa 0.5 micron 15 ml ultra centrifugation filter and washed 3x with 10 ml of Nanopure $H_2O$.

**Probe Characterization:**

Transmission electron microscope (TEM) images were collected on a Phillips CM-120 transmission electron microscope operating at 80keV. A minimum of 500 particles were measured to determine particle diameter. Dynamic light scattering (DLS) was used to determine the average hydrodynamic particle size in solution with a Nanotrac 150 particle size analyzer (Microtrac, Inc., Montgomeryville, PA) using a 150s scan time fitted with a residual less geometric eight-root regression. Elemental analysis for copper and iron was carried out with a Varian AA 220FS operated under a flow of air/acetylene and a copper lamp set to 248.3 nm and 324.8 nm for iron and copper, respectively. Relaxivity values were measured using a Bruker Minispec mq60 relaxometer operating at 1.4T, and 37 $^{o}$C. Dextran coating was verified by infrared (IR) spectroscopy performed on a Shimadzu IR Prestige 21 Spectrophotometer using diffuse diffraction with KBr as the dilutant. Two tailed student T tests with equal variances of n=3 was performed to test the significance between DIO/Cu points and DIO vs DIO/Cu points at the 95% confidence levels. $^{64}$Cu levels were monitored with a Fluke Biomedical CAL/RAD Mark IV Dose Calibrator and inductively coupled plasma-mass spectrometry (ICP-MS) was also performed. X-ray diffraction patterns were taken on a Siemens D-500 powder diffractometer with a Cu source. Room temperature magnetometry studies were performed on a Princeton Measurement Corporation vibrating sample magnetometer (VSM), while temperature dependent and cryogenic measurements were performed on a Quantum Design superconducting quantum interference device (SQUID) magnetometer.

**Results and Discussion**

Hydrodynamic radius of the microwave-assisted synthesized nanoparticles were measured using DLS. Using a maximum power of 300W (n=3) a linear trend can be observed between particle size and increased reaction time. Products were shown to increase in size by roughly 10 nm for every 5 minutes of reaction time as shown in Figure 1(a). In traditional methods, 2 h of reflux are required to produce similar results. The microwave-assisted technique is able to reduce reaction times by providing uniform heating caused by an increase in turbid motion when the microwave rays enter the sample. This rapid motion causes an instantaneous heating at the molecular level opposed to traditional conduction methods, such as refluxing. Rapid and uniform heating allows for a more uniform and controllable size distribution.

Conversely, a decreasing trend in size was seen when the maximum power was increased as shown in Figure 1(b). All produced particles were within the optimal size range for *in vivo* testing (d<100nm), however, particles produced at 300 W were the smallest and of greatest interest to medical applications, so these were carried forward for further characterization.

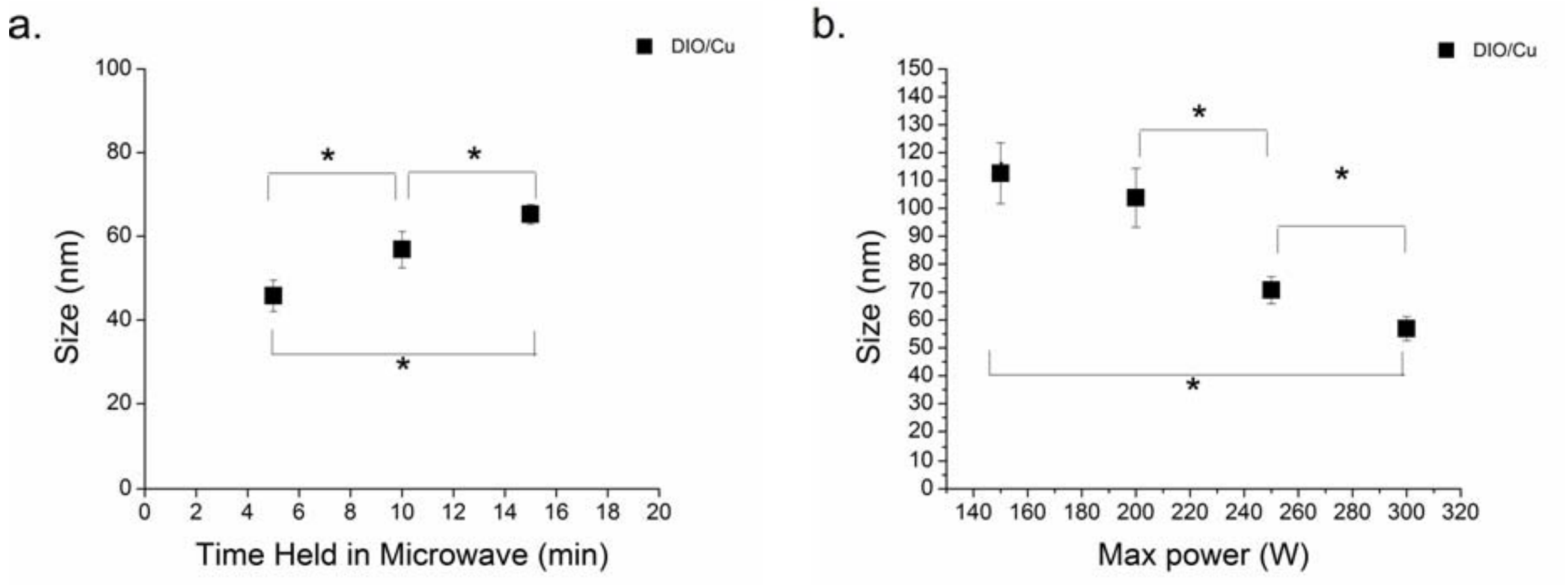

**Figure 1.** Dependence of particle size on duration of heating and max power. (a) Size increased with duration of heating time in the microwave while the max power was set to 300 W (DIO/Cu = ■) (b) Size decreased when max power was increased and time was set to 10 minutes, as shown for (DIO/Cu= ■). [*] designates the statistical significance between different trials where $p<0.05$

Atomic absorption data, Figure 2, suggests that there was no significant difference in the fractional iron and copper content in the products from the different reaction times, regardless of hydrodynamic size. This suggests that after a brief nucleation period, dextran encapsulates the iron oxide particle, halting further growth. Further reaction time tends to increase the size of the particle both by adhering new DIO particles and by adding additional dextran. This allows for an increase in hydrodynamic radius while fractional iron content remains the same within the particle. Copper content was noted to remain the same within the different reaction conditions. However, there was a significant difference in copper quantities found in DIO/Cu vs the control group DIO, signifying the successful incorporation of copper. Micromolar quantities of copper were incorporated within the iron oxide. This far exceeds the range of concentration required for PET imaging use, as PET can be sensitive from nano to picomolar quantities *in vivo.*[19] An incorporation efficiency of 10% was achieved, which is higher than previously reported efficiencies of 5.2% for chelation of copper to nanoparticle surfaces.[7]

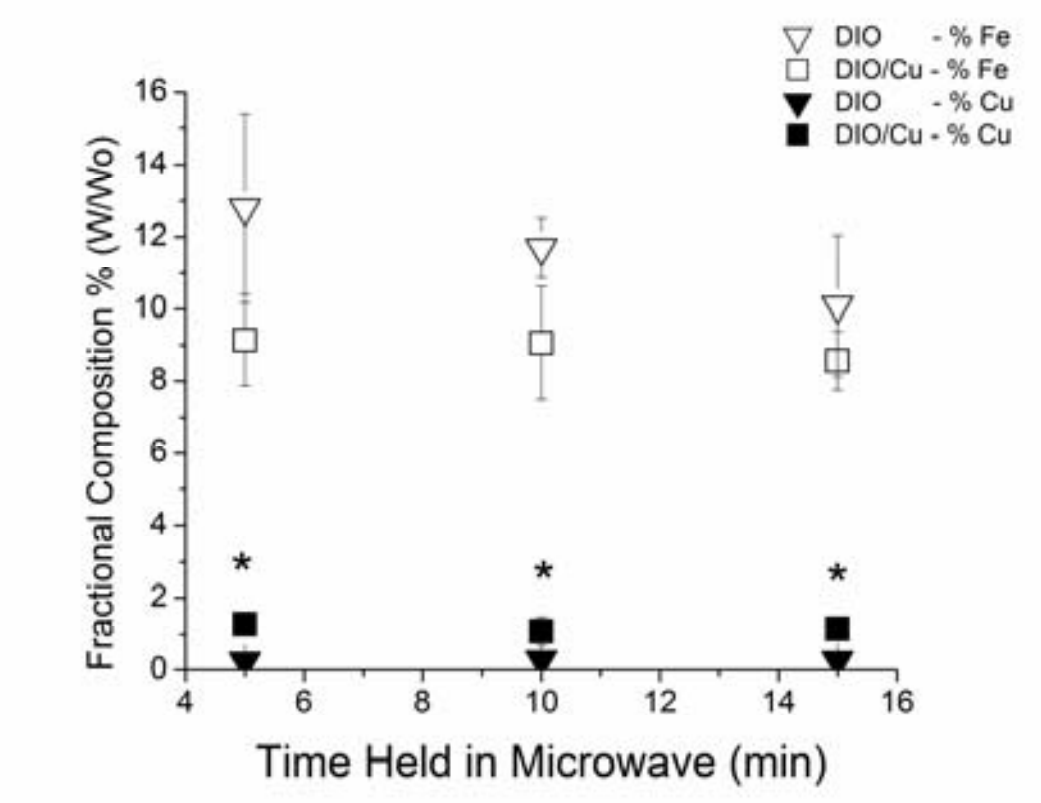

**Figure 2. Iron and Copper Content for the Microwave Synthesized DIO and DIO/Cu.** Percentage of iron (Open) and copper (Closed) is shown where ▼ is representative of DIO, copper less reaction and ■ as DIO/Cu. Iron content in DIO can be seen to be slightly higher than that of DIO/Cu demonstrating that the incorporation of copper affects the iron content within the nanoparticle. Increase of copper levels seen in DIO/Cu compared to DIO was measured to be a 1% doping level of copper W/W%. A significant change in iron content was not detected despite size increases over time signifying that the core composition remains the same leaving changes only in the shell size. [*] designates the statistical significance between doped and undoped trials where $p<0.05$

Relaxometry was performed on the samples and is shown in Figure 3. For DIO/Cu there was no significant change in $r_1$ (panel a) and $r_2$ (panel b) values regardless of overall (DLS) diameter. Relaxivity values of DIO and DIO/Cu were not significantly different between products from different reaction times. This confirms our earlier suggestion that the core particles form for only a short period of time and further suggests that the iron oxide cores within a larger particle are weakly interacting. However, relaxivities were significantly higher in the copper-doped DIO compared to the control DIO. In general the copper doped material was of smaller total size, contained significantly more copper, but iron content was not significantly different ($p < 0.05$).

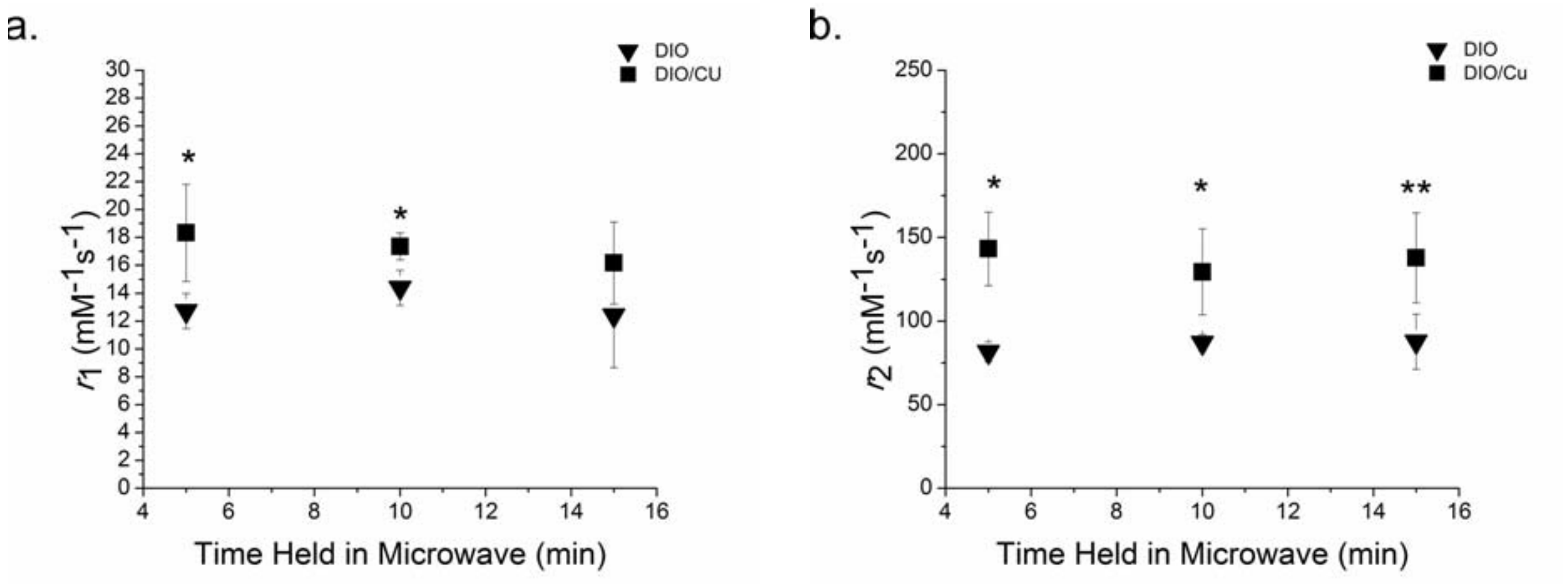

**Figure 3. Relaxivity values for the microwave synthesized DIO and DIO/Cu.** Relaxivity, $r_1$ (a) and $r_2$ (b). As anticipated, there was no significant difference between $r_1$ and $r_2$ values between different timed reactions due to equal iron concentrations within the nanoparticle products. However, a significant difference was seen between DIO/Cu = ■, DIO =▼. [*, **] designates the statistical significance between different trials where *$p<0.05$, **$p<0.10$)

AA, DLS and relaxivity results are summarized in Table 1; these data suggest that the increased particle size that occurs under prolonged microwave reaction times is strictly due to increased thickness of surface coating and accumulation of iron cores, not an increase in the size of the cores; this is evident by the lack of change in iron content and relaxivity with synthesis time. The relaxivity values shown for DIO/Cu are comparable to commercial agent such as Feridex© and Resovist© which have longitudinal relaxivity ($r_1$) of 12.3 and 25 $mM^{-1}$ $s^{-1}$and transverse relaxivity ($r_2$) of 191[20] and 151[21] $mM^{-1}$ $s^{-1}$, respectively.

| Type | Time (min) | Total Size (nm) | Core Size (nm) | Fe (mol) | Cu (mol) | $r_1$ ($mM^{-1}s^{-1}$) | $r_2$ ($mM^{-1}s^{-1}$) |
|---|---|---|---|---|---|---|---|
| DIO/Cu | 5 | 45.7 ± 3.7 | 5.3 ± 1.8 | $1.6E^{-04} \pm 2.1E^{-05}$ | $1.3E^{-05} \pm 1.0E^{-05}$ | 17.1 ± 2.3 | 135.8 ± 30.9 |
| DIO | 5 | 61.9 ± 12.6 | 3.3 ± 1.3 | $1.1E^{-04} \pm 2.1E^{-05}$ | $1.8E^{-06} \pm 3.4E^{-07}$ | 15.1 ± 2.3 | 88.2 ± 8.8 |
| DIO/Cu | 10 | 56.8 ± 4.3 | 3.1 ± 0.9 | $1.2E^{-04} \pm 4.3E^{-05}$ | $7.7E^{-06} \pm 7.0E^{-06}$ | 15.8 ± 3.4 | 128.1 ± 27.9 |
| DIO | 10 | 63.9 ± 5.8 | 3.4 ± 1.6 | $9.6E^{-05} \pm 7.3E^{-06}$ | $2.1E^{-06} \pm 2.0E^{-07}$ | 14.9 ± 3.8 | 93.4 ± 23.3 |
| DIO/Cu | 15 | 65.3 ± 2.4 | 3.3 ± 0.9 | $1.5E^{-04} \pm 1.1E^{-05}$ | $9.6E^{-06} \pm 9.3E^{-06}$ | 15.5 ± 1.5 | 126.5 ± 31.2 |
| DIO | 15 | 71.8 ± 16.6 | 3.1 ± 1.4 | $9.6E^{-05} \pm 2.5E^{-05}$ | $2.2E^{-06} \pm 5.0E^{-07}$ | 11.6 ± 4.9 | 79.5 ± 28.0 |

**Table 1**. Table comparing DIO and DIO/Cu trends for total size, core size, moles of iron, moles of copper, $r_1$ and $r_2$ values.

Magnetometry of DIO and DIO/Cu nanoparticles was performed; room temperature magnetic hysteresis loops are shown in Figure 4. Both the DIO and DIO/Cu were found to be superparamagnetic at room temperature, a necessity for application to MRI, with a blocking temperature of ~40 K. From the blocking temperature we estimate the size of the doped and undoped iron oxide cores[22] to be about 8 nm. Saturation magnetization was measured at 10 K and found to be 16 emu/g and 12 emu/g for the DIO and DIO/Cu, respectively, normalized to the mass of the entire particle including the dextran coating. This reduction in saturation magnetization in DIO/Cu is consistent with the inclusion of Cu in exchange for Fe.

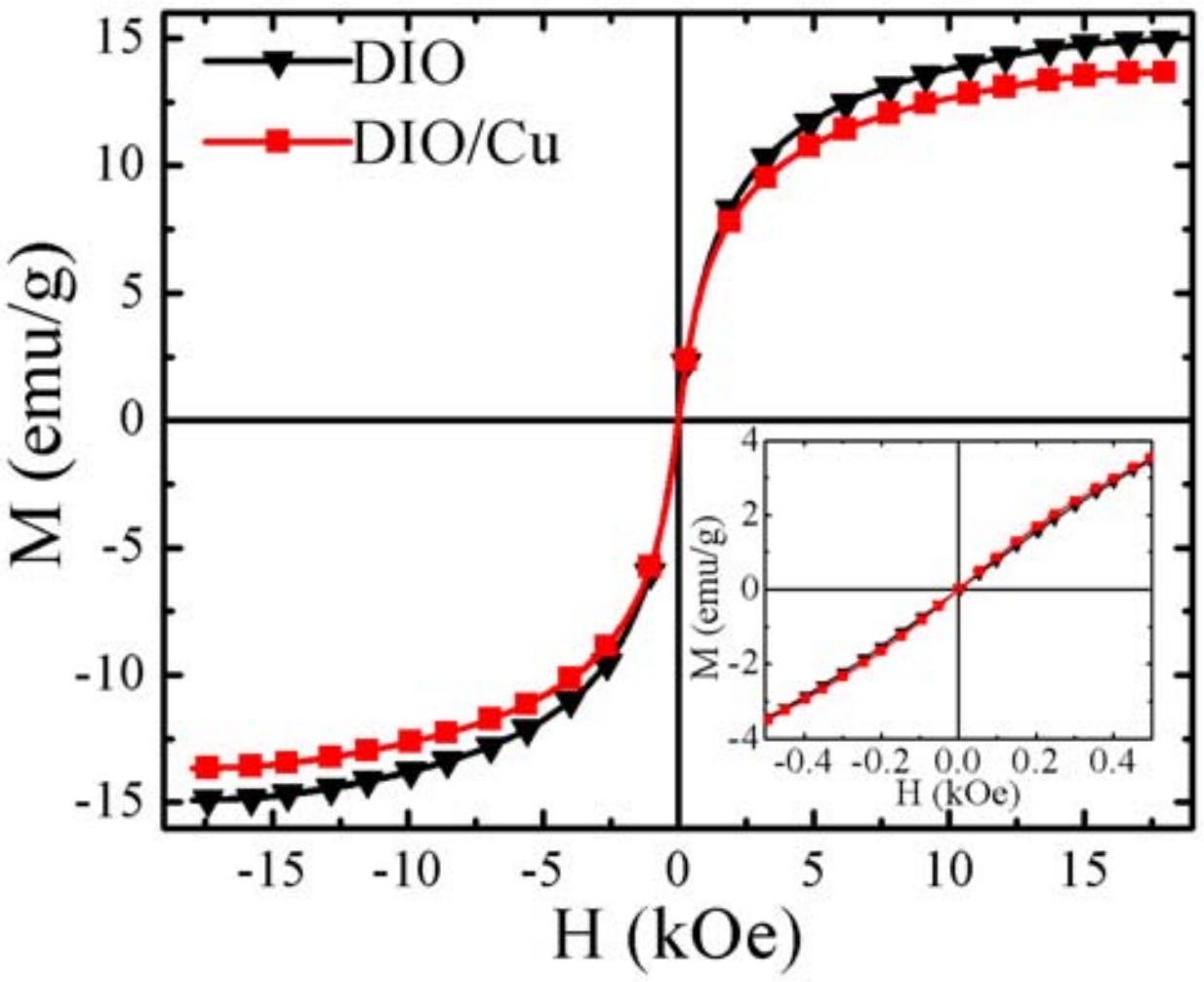

**Figure 4. Hysteresis of DIO and Dio/Cu Nanoparticles.** Room temperature magnetic hysteresis loops of DIO and DIO/Cu nanoparticles. Inset shows a zoom-in view, confirming that these particles are superparamagnetic.

Dextran coating of iron oxide doped with copper allows for increased biocompatibility and blood retention time. The addition of dextran also aids in size control; in the absence of dextran, bare iron oxide nanoparticles tend to aggregate as shown in Figure 5(a) with a core size of 11.6 ±2.9nm. Iron oxide particles synthesized in the presence of dextran (DIO, Figure 5b) were more dispersed and core sizes decreased to 3-6nm. This was also observed for DIO/Cu. It can be noted that there were multiple 3 nm cores per 50 nm particles. The similarity in core sizes indicates that the change in overall hydrodynamic size is due to thicker dextran coating.

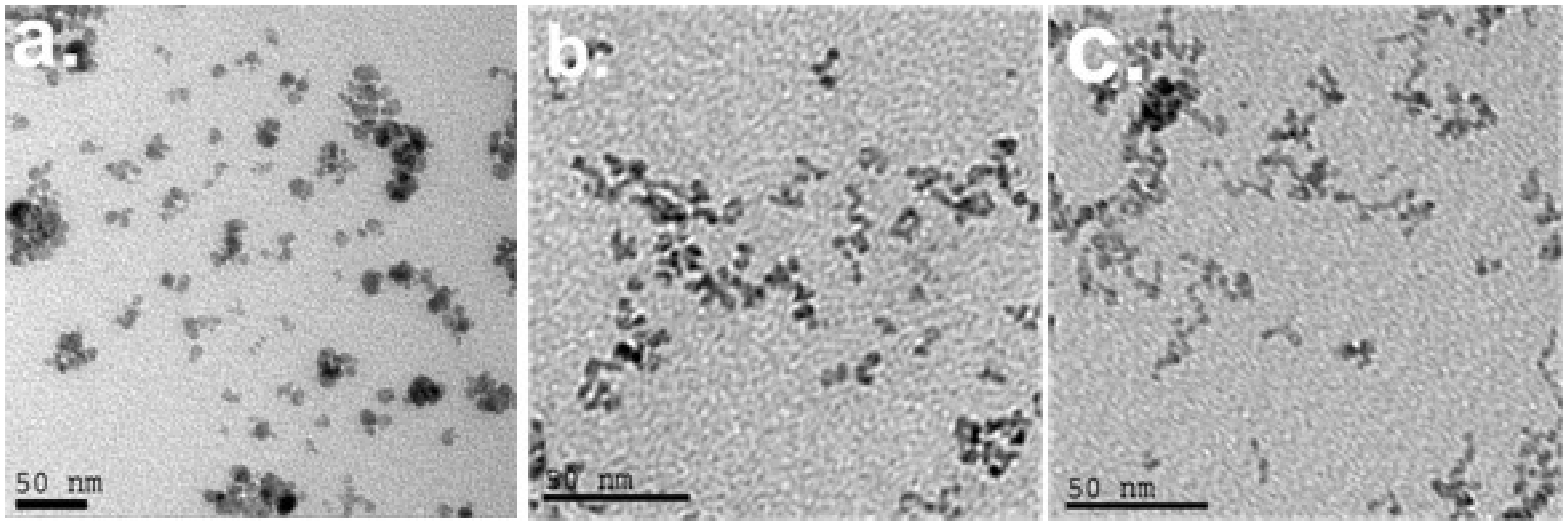

**Figure 5. TEM of a. bare iron oxide, b. DIO and c. DIO/Cu Nanoparticles.** Bare iron oxide prepared using a similar protocol with no dextran present (a) shows iron oxide aggregating with increased core sizes. Nanoparticles prepared in the presence of dextran (B and c) are significantly more dispersed in nanoparticles containing copper and those without.

As further evidence of dextran coating FTIR was performed and is shown in Figure 6. Dextran vibration peaks (Figure 6a) include the C-O stretch at 1015 $cm^{-1}$, C-H bend at 1350$cm^{-1}$, C-H stretch at 2900$cm^{-1}$ and O-H stretch at 3350$cm^{-1}$. These same peaks are found for DIO (Figure 5c) and DIO/Cu (Figure 5d) but not bare iron oxide particles (Figure 6b). All of the particles showed peaks characteristic of hydroxyls, i.e. the O-H bend at 1404$cm^{-1}$ and O-H stretches at 3410$cm^{-1}$.The hydroxyl peaks observed may be attributed to water. In combination with the increased size and dispersion illustrated by TEM in Figure 5, the FTIR results further confirm the coating of dextran onto the DIO and DIO/Cu nanoparticles.

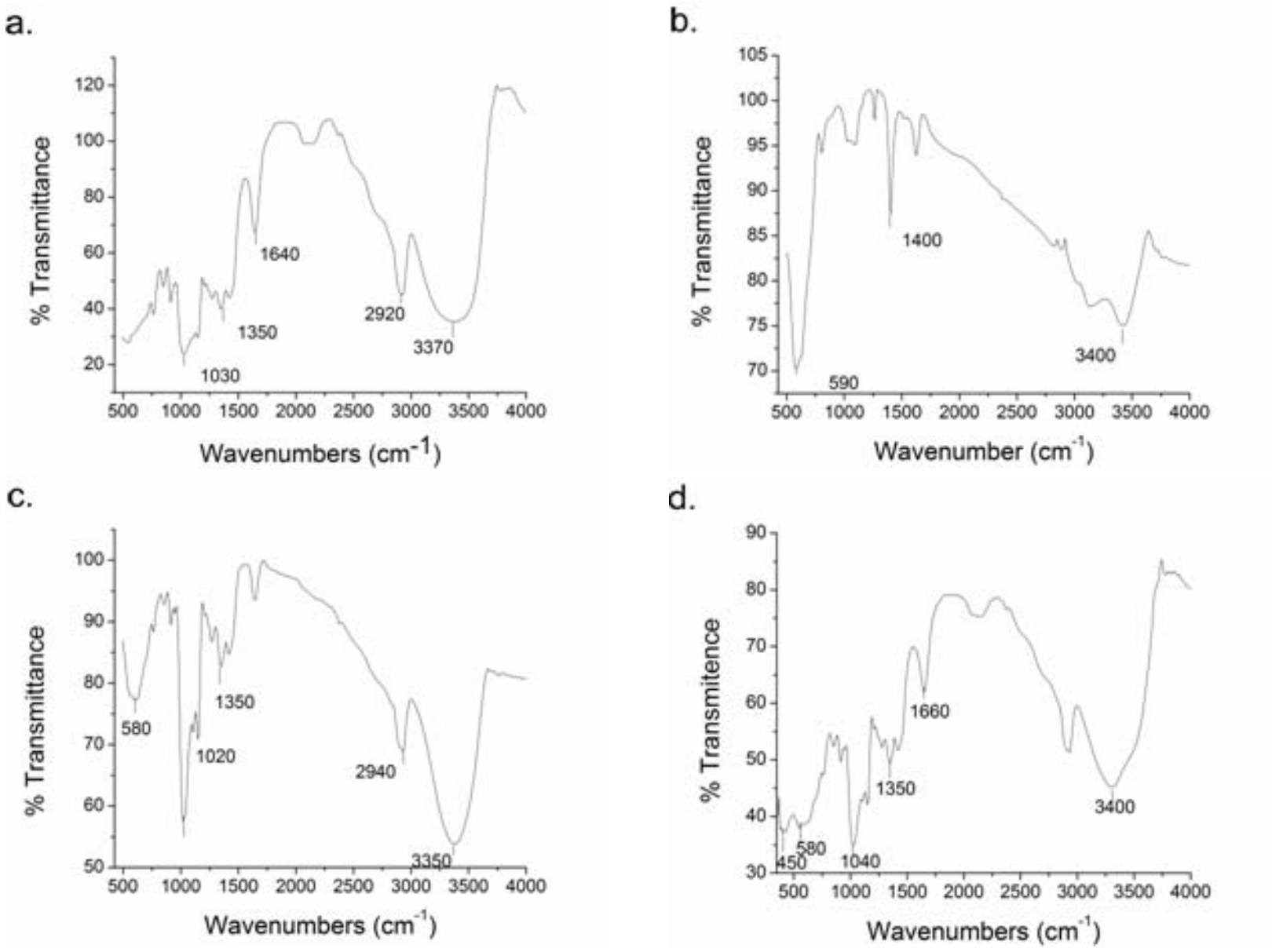

**Figure 6.** FTIR of a) dextran, b) bare iron oxide, c) DIO, D) DIO/Cu

Powder X-ray diffraction was performed to determine the structural effect of the incorporation of copper as well as the structure of iron oxide. Diffraction pattern for DIO is shown in Figure 7a and indicate the dominant phases are magnetite ($Fe_3O_4$) and/or maghemite ($\gamma$- $Fe_2O_3$); the presence of $Fe^{2+}$ in the synthesis process does favor the magnetite phase. Copper added during copreciation is in the 2+ oxidation state, working in direct competition with $Fe^{2+}$ (from $FeCl_2$) to occupy the octahedral site in the $Fe_3O_4$ structure, forming the cuprospinel structure. Due to the small amount of copper doping (10%) the patterns for b) DIO/Cu, and c) DIO/$^{64}$Cu do not change signifigantly compared to a) undoped DIO. A small peak at $2\theta=74.5^{o}$ in the DIO/Cu sample suggests the formation of the cuprospinel structure, confirming the $Cu^{2+}$ replacement of the $Fe^{2+}$ in the original magnetite phase.

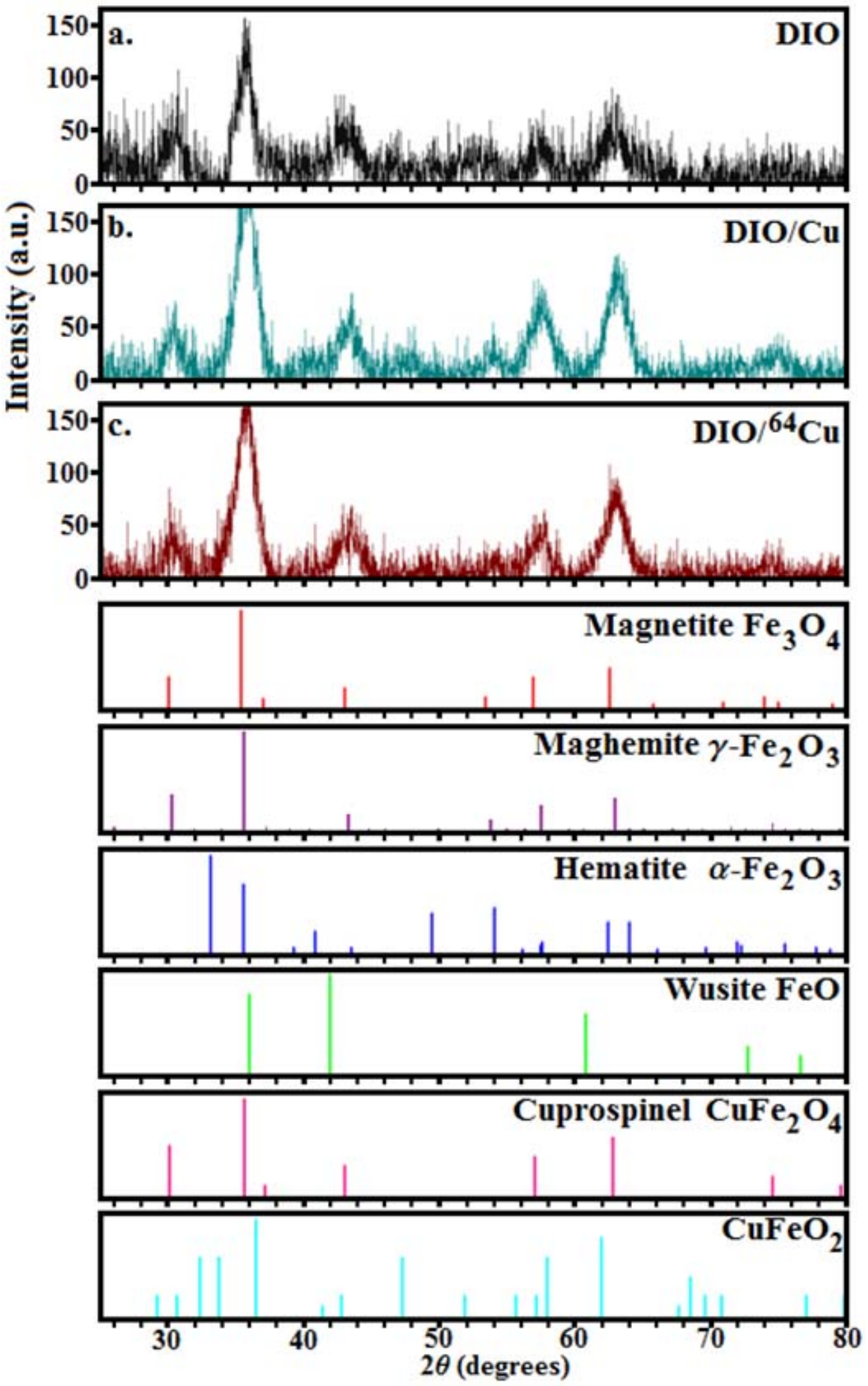

**Figure 7. X-Ray Diffraction for Nanoparticles.** X-ray diffraction patterns for (a) DIO, (b) DIO/Cu, and (c) DIO/$^{64}$Cu and reference patterns for Magnetite, Maghemite, Hematite, Wusite,

**DIO/Cu-64 Synthesis**

Preliminary radiolabeling studies were performed to incorporate of copper-64 into DIO/Cu nanoparticles. These experiments were performed on the same make and model of microwave used for the above studies but on a different machine that was approved for radioactive use. The primary differences for the instrument used for radiolabeling were that it has a higher cooling air pressure and less temperature fluctuations. To control for instrument dependent artifacts DIO, and DIO/Cu-64 were synthesized on this new instrument using a 1 minute 300W synthesis. As shown on Table 2, the hydrodynamic sizes were noted to be 36 ± 2, and 35 ± 3 nm respectively as shown in supplemental data (S1). Trends demonstrating size increases with increased time were also noted with the new instrument (data not shown). $R_2$ values were measured to be to be 121 ± 14, and 100 ± 20 $mM^{-1}s^{-1}$ respectively. Using TEM, core sizes were observed to be 5.4 ± 2.3, 4.2 ± 1.9 nm (S2) respectively. The XRD pattern for the DIO/$^{64}$Cu is identical to DIO/Cu and both are similar to DIO.

| Type | Total Size (nm) | Core Size (nm) | $r_2$ ($mM^{-1}s^{-1}$) |
|---|---|---|---|
| DIO | 35.6 ± 1.5, | 5.4 ± 2.3 | 121 ± 14 |
| DIO/$^{64}$Cu | 35.3 ± 3.3 | 4.2 ± 1.9 | 100 ± 20 |

**Table 2**. Table comparing DIO and DIO/$^{64}$Cu trends for total size, core size and $r_2$ values for radio labeling conditions.

The presence of copper-64 was verified with gamma counting after purification to remove free, unincorporated copper. Initially 720 ± 20 µCi of $^{64}$Cu was added and after purification (with decay correction) 218 ± 19 µCi remained, representing a 33% incorporation. The use of a high powered microwave for radiolabeling has a higher incorporation efficiency of Cu-64 compared to other previously mentioned radiolabeling methods[7] as was observed for cold copper. ICP-MS was also performed and demonstrated that the elemental content in DIO/$^{64}$Cu for iron and copper were 1.1$E^{-2}$ mg/ml and 9.7$E^{-4}$ mg respectively. This corresponds to roughly 4.9$E^{-11}$ ± 1.4$E^{-12}$ moles of copper per sample. It is important to note that the sensitivity of PET is between nanomolar and picomolar levels making this probe well within the range of detectibility[23]. Furthermore, the starting amount of $^{64}$Cu used for this preliminary study was far less than the amounts of cold copper used in the proof-of-concept studies (5.3$E^{-12}$ ± 5.5$E^{-13}$ moles of $^{64}$Cu, compared to 1.5$E^{-5}$ moles of cold copper). Higher $^{64}$Cu content could be conceivably achieved by increasing the amount of $^{64}$Cu initially included in the reaction.

**Conclusion**

In summary, we have demonstrated a microwave assisted synthesis technique to produce copper-doped and $^{64}$Cu-doped, iron oxide nanoparticles with a dextran coating. This novel synthesis allows for the rapid, size-controlled production of DIO/Cu, a novel contrast agent that is a proof-of-principle model for dual probe PET/MRI probes. Through the use of microwave heating, reaction times were significantly decreased from 120 minutes to 5 minutes while yielding similar sized particles. Size of the nanoparticles can be controlled simply by increasing the microwave heating time or by decreasing the maximum power. The rapid synthesis is greatly beneficial by leaving the radiolabel less time to decay.

The size distribution of DIO/Cu shows that the synthesized core-shell nanoparticles can be achieved with diameters between 20 to 100nm, which is in a desirable size range. Size controlled synthesis of iron oxide + dextran particles is necessary due to the dependence of the magnetic states on the dispersion and aggregation, and the influence of size on the blood retention time. Nanoparticles over 80 nm are quickly removed by the reticuloendothelial system while nanoparticles smaller than 7-8nm are removed by the kidneys,[24, 25] leaving a finite range in which nanoparticles have optimal blood retention time

FTIR and TEM support that dextran is coated onto the iron oxide. The coating of dextran increases biocompatibility and allows for the ability for surface functionalization with specific targeting of biomarkers for disease[26]. Syntheses using $^{64}$Cu were able to incorporate microCi amounts of Cu-64 into the iron oxide particles, which is in the range of sensitivity for PET imaging. The relaxivities show consistent $r_1$ and $r_2$ values despite changes in overall size and the values obtained were comparable to commercially available standard probes when incorporated with copper. Further surface modifications will be done to allow for key targeting of areas of interest as well as *in vitro* and *in vivo* studies.

**Acknowledgements**

This work has been supported by the Department of Energy (DOE DESC0002289) the National Science Foundation (DMR-1008791).

## Supplemental Data

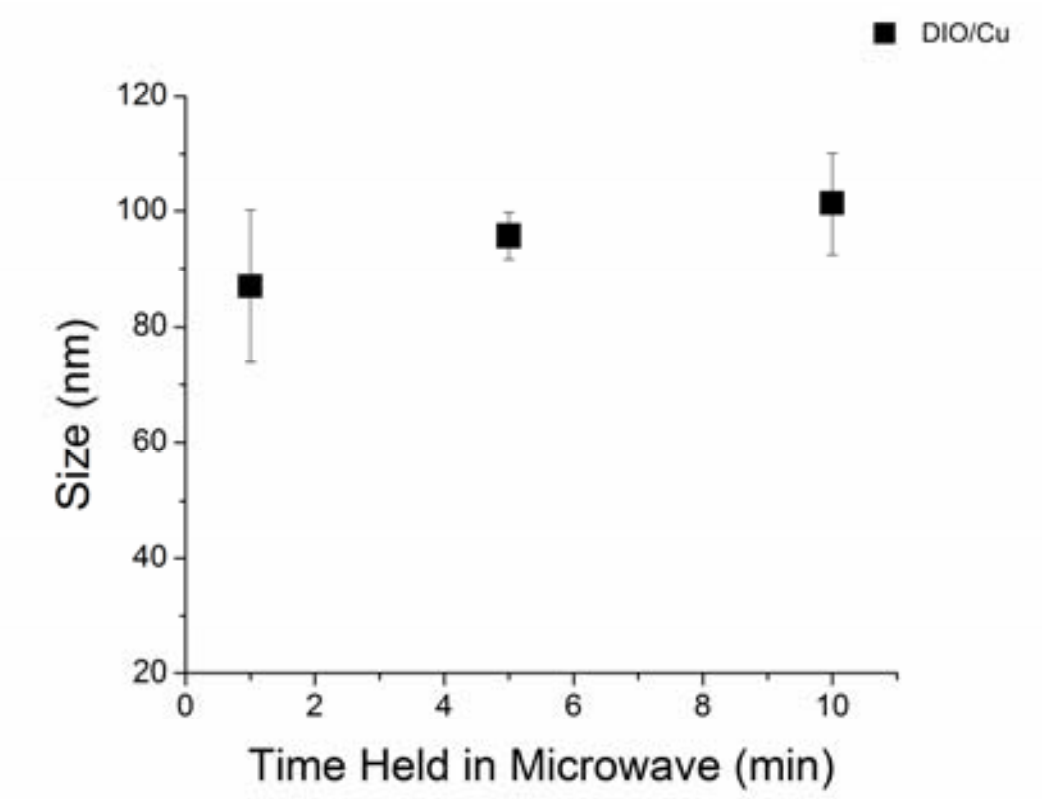

**S1. DLS of Radiolabel conditions.** DLS of 1 minute reactions at 85 $^{o}$C with a max power of 300 W on a) DIO, b) DIO/Cu, c) DIO/Cu-64. Average hydrodynamic sizes for particles were 35.6 ± 1.5, 59.6 ± 14.19, and 35.3 ± 3.3 nm respectively with P value < 0.1 between all sets of data.

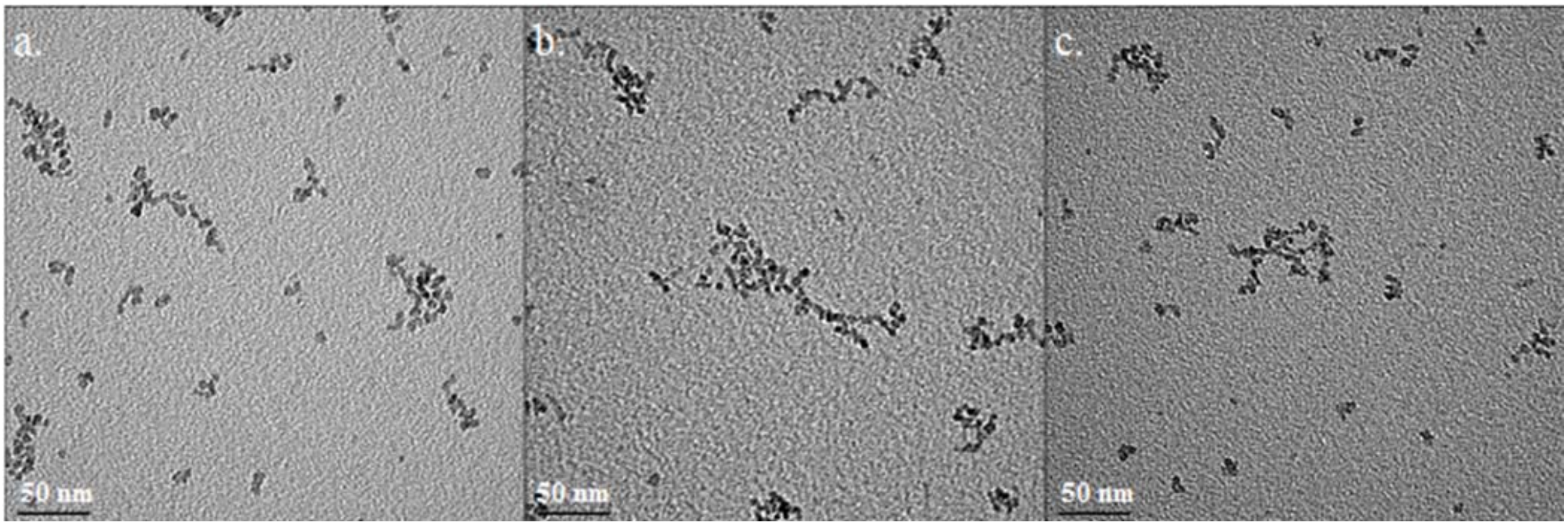

**S2. TEM of Radiolabel Conditions.** TEM of 1 minute reactions at 85 $^{o}$C with a max power of 300 W on a) DIO, b) DIO/Cu, c) DIO/Cu-64. With n=500, core sizes were noted to be a) 5.4 ± 2.3 nm, b) 4.6 ± 1.7 nm, c) 4.2 ± 1.9 nm.